\documentclass[pageno]{jpaper}

\usepackage{amsmath,amssymb,amsfonts, amsthm}
\usepackage[ruled, linesnumbered]{algorithm2e}
\usepackage{pifont}
\usepackage{enumitem}

\usepackage{algorithmic}
\usepackage[normalem]{ulem}

\newcommand{\name}{\textsc{Metro}}

\author{
Zhao Wang \\
Peking University \\
\texttt{wangzhao21@pku.edu.cn}
\and
Jingchen Zhu \\
Peking University \\
\texttt{zjc990112@pku.edu.cn}
\and
Zhe Zhou \\
Peking University \\
\texttt{zhou.zhe@pku.edu.cn}
\and 
Guangyu Sun \\
Peking University \\
\texttt{GSun@pku.edu.cn}
}

\begin{document}

\title{\textsc{Metro}: a Win-Win Solution to Provide Efficient Communication on Spatial Accelerators for Tensor Applications}

\date{}
\maketitle

\thispagestyle{empty}

\begin{abstract}
Spatial architecture is an effective solution to build  large-scale accelerators for tensor applications. As more and more cores are integrated on a spatial accelerator, the interconnection becomes a critical design factor. However, the traditional NoC architecture cannot handle the seriously unbalanced traffic of processing tensor applications on a spatial accelerator. We find that its inefficiency is fundamentally caused by hardware-based traffic scheduling design, which has intrinsic limitations. To overcome the limitations, we propose to decouple the traffic scheduling from hardware fabrics and move it to the software level for a win-win solution. First, the software has a global view of the traffic to find better scheduling strategies than hardware-based ones. Second, after offloading the traffic control to software, the hardware design of NoC can be simplified. To achieve this goal, we propose a holistic solution called \name, which consists of a software framework and a dedicated hardware design.  We evaluate the efficiency of \name\ with widely-used  tensor  applications. Experimental results prove its  advantages  over traditional hardware design,  in  respect  of  performance, energy consumption, and hardware design complexity.

\end{abstract}

\section{Introduction}


Tensor applications, which involve intensive tensor computation, have dominated in many important domains, including computer vision, natural language processing, data mining, etc~\cite{resnet, gpt, bert, multilinear, tensorda, efficientspmv}. Recently, researchers have proposed many dedicated accelerators to improve the processing efficiency of tensor applications~\cite{eyeriss, eyerissv2, tpuv4, simba, graphcore}. Among these approaches, the spatial accelerator is demonstrated to be a promising solution to building a high-performance and scalable processing system for these applications.

A typical spatial accelerator consists of an array of processing cores, a memory hierarchy, and the interconnection for communication. Recently proposed spatial accelerators keep increasing the number of cores to meet the computation requirements of tensor applications. For example, a package of Simba contains 576 cores providing 32~TOPs computation throughput~\cite{simba}; The wormhole processor from Tenstorrent achieves 430~TOPS with 80 customized cores~\cite{tens-wormhole}; Graphcore IPU integrates 1472 computing engines with over 8000 parallel threads~\cite{graphcore}. 


To exploit the massive cores, the interconnection becomes a critical design factor. State-of-the-art spatial accelerators employ the network-on-chip~(NoC) architecture to provide high-performance and flexible interconnection~\cite{tangram, simba, graphcore, eyerissv2, maeri}. However, it is challenging for a traditional NoC design to handle the traffic of processing tensor applications. The traffic is dominated by bursty, large-size, and collective data communication, which results in seriously unbalanced utilization of NoC. There exist many works for collective communication in a NoC design~\cite{bam, rpm, whirl}. However, they cannot achieve high-performance and high utilization at the same time, when processing tensor applications on a spatial accelerator. 

The inefficiency of a traditional NoC architecture is fundamentally caused by a hardware-based traffic scheduling design, which has intrinsic limitations. First, due to the area and timing constraints, a hardware-based traffic scheduling strategy is normally simple and rigid. Second, a router can only sense the traffic status around it and decide by itself. Thus, it cannot co-operate with other routers to optimize its traffic scheduling strategy according to the global status. Consequently, although there exist opportunities to optimize traffic scheduling, it is impossible for a hardware-based design to find and leverage them.

To overcome the limitations, we propose to remove the traffic scheduling from the hardware and move it \textbf{entirely} to the software level. The rationale is that the software schedules the tensor application onto the spatial accelerator for computation. Thus, the software can also analyze the communication information in advance and find proper traffic scheduling strategies. The benefits are two-folded. First, software has the global view of the traffic to find better scheduling policies than hardware scheduling can do. Second, after offloading the traffic control to software, the hardware design of NoC can be simplified. 

Two important problems should be solved to enable software-based traffic scheduling. First, given a tensor application and a spatial accelerator, we need to formulate the design space of traffic scheduling strategies from the software perspective. In addition, we need to provide an efficient method to find a proper solution in the large design space. Second, the NoC hardware should be modified to support the software-based traffic scheduling. We find that a software traffic scheduling strategy may require hundreds or even thousands of bits to store the path routing and flow control information. Thus, the techniques are needed to compress such information and reduce overhead.  


\begin{figure*}[t]
  \centering
  \includegraphics[width=\textwidth]{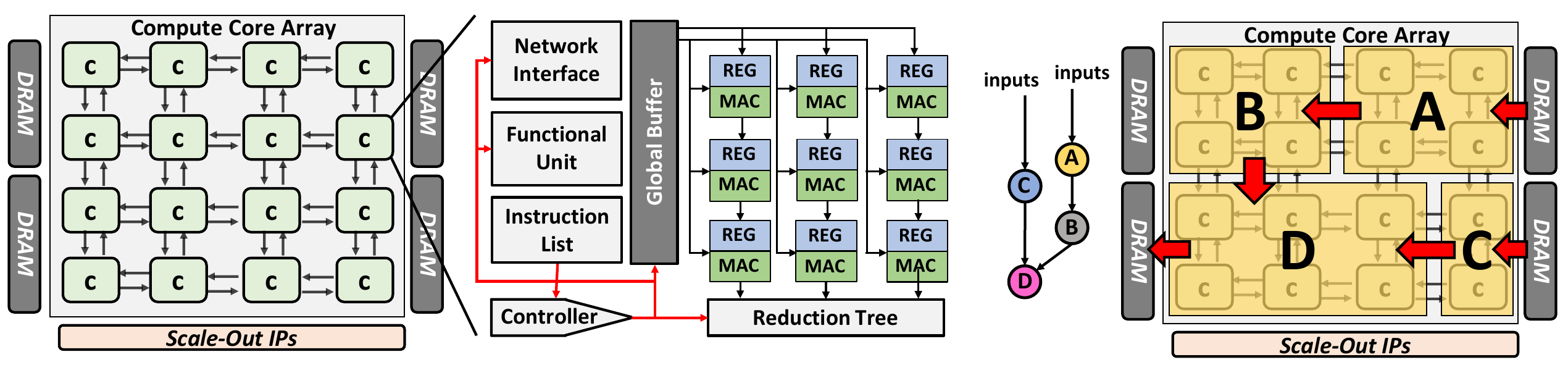}
  \caption{(a) A spatial accelerator. (b) A typical computing core. (c) Mapping a tensor operator DAG onto the accelerator. } 
  \label{fig:sa-basic}
\end{figure*}
To address these problems, we propose a holistic solution called \name, which consists of a software scheduling framework and a customized NoC hardware. The name is inspired by the real metro traffic, which is also well scheduled in advance before operation. The drivers do not need to worry about path routing and traffic control during operation. The main contributions of this work are summarized as follows, 
\begin{itemize}

\item We find that the traditional hardware-based traffic scheduling has intrinsic limitations to handle the seriously unbalanced traffic of processing tensor applications on a spatial accelerator. 

\item By leveraging the prior knowledge of communication, we propose a win-win solution to offload the traffic scheduling policies to the software level. It not only mitigates the problem of unbalanced traffic but also can help reduce the hardware design complexity. 

\item We formulate the traffic scheduling strategy and construct its design space. We provide a method to explore the design space and find the proper strategy. We propose a method to control the traffic injection time so that the traffic conflicts can be further reduced. 


\item We propose a dedicated NoC hardware design to support the software based traffic scheduling. We propose a hybrid routing mechanism, which can reduce the information length to store the strategy.

\item We provide comprehensive experimental results to prove the advantages of \name, in respect of performance, energy consumption, utilization, hardware design complexity, etc. 


\end{itemize}


\section{Background and Motivation}






\subsection{Spatial Accelerators for Tensor Applications} \label{subsec:spatial_arch}
The spatial architecture normally refers to a category of designs that employ an array of processing cores to exploit the high processing parallelism of applications. Figure~\ref{fig:sa-basic}~(a) illustrates a typical spatial accelerator for tensor applications. Sixteen cores are connected with a 2D mesh NoC. These cores can access the DRAM with four memory controllers. Note that it also provides a scale-out interface, which can further extend its computation capability by connecting with one or more spatial accelerators. 

As shown in Figure~\ref{fig:sa-basic}~(b), each core consists of multiple MAC units, which can conduct multiply-accumulate operations in parallel. The partial sums of MACs are further reduced by a reduction tree. The core also integrates a functional unit to perform complex mathematical operators, such as sigmod and pooling. It utilizes an on-chip network interface to packetize and un-packetize packets. These components share a global scratched buffer to exchange intermediate data. A controller harmonizes these hardware components according to instructions. 


When processing a tensor operator in a core, the MAC units are normally organized with a specific dataflow. The term dataflow refers to the location and sequence of processing a tensor operator on these MAC units~\cite{tenet, maeri, heterogeneous, maestro, interstellar}. Various dataflows have been proposed. For example, TPU employs a systolic-array dataflow in each core~\cite{tpu}; Simba~\cite{simba} employs a weight-stationary dataflow inside a core;  NN-Baton~\cite{nn_baton} uses an output-stationary style for each core.

A large tensor application is normally composed of multiple tensor operators organized as a DAG, as shown in Figure~\ref{fig:sa-basic}~(c). Thus, when processing a tensor application on a spatial accelerator, we need to {\bf map} these operators on different processing cores. This is also illustrated in Figure~\ref{fig:sa-basic}~(d). To reduce the cost of accessing off-chip memory, these operators forwards intermediate data between operators through the on-chip network instead of storing them back to the off-chip memory.

Recently, there is a strong interest in scaling up the performance of spatial accelerators~\cite{tangram, atomic, multi_nn, heterogeneous, simba, scaledeep}. First, tensor applications increase rapidly in computing intensity, memory footprint, and operator graph complexity~\cite{bert, gpt}. Second, the development of manufacturing technologies, such as 2.5D/3D stacking~\cite{tetris} and chiplet integration~\cite{simba, nn_baton}, have enabled more and more hardware resources on a spatial accelerator.

\subsection{Traffic of Processing Tensor Applications}~\label{subsec:traffic}
The traffic across the NoC is seriously unbalanced when processing tensor applications. The situation is caused by two intrinsic traffic features.

First, spatial hotspots widely exist due to the heavy multicast and reduction traffic. Paralleling a tensor application typically relies on collective communication to carry data across computing cores. For example, training a DNN model in data-parallel updates gradients with all-reduce traffic, which is typically implemented with the combination of multicast and reduction patterns. These patterns aggregate various messages from~(to)~the source~(destination)~nodes. And the messages compete for nearby channels and lead to traffic hotspots, which dramatically increase the congestion probability.



Second, the bursty traffic exacerbates the effect of hotspot congestion. Bursty traffic is due to the large message size and high injection sparsity. The tensor processing algorithms prefer long messages to maximize data reuse within a core. 
It takes hundreds to thousands of cycles for a long message to pass over a channel. Once congestion happens, the transmission latency of a blocked message increases dramatically due to waiting for the channel to free. Moreover, most tensor operations exhibit high computing intensity. Thus, the time interval between two bursty traffic is also long. It means that the message injection demonstrates high temporal sparsity.

\subsection{Limitations of Existing Solutions}  \label{subsec:candidate}
\begin{figure}[ht]
  \centering
  \includegraphics[width=0.48\textwidth]{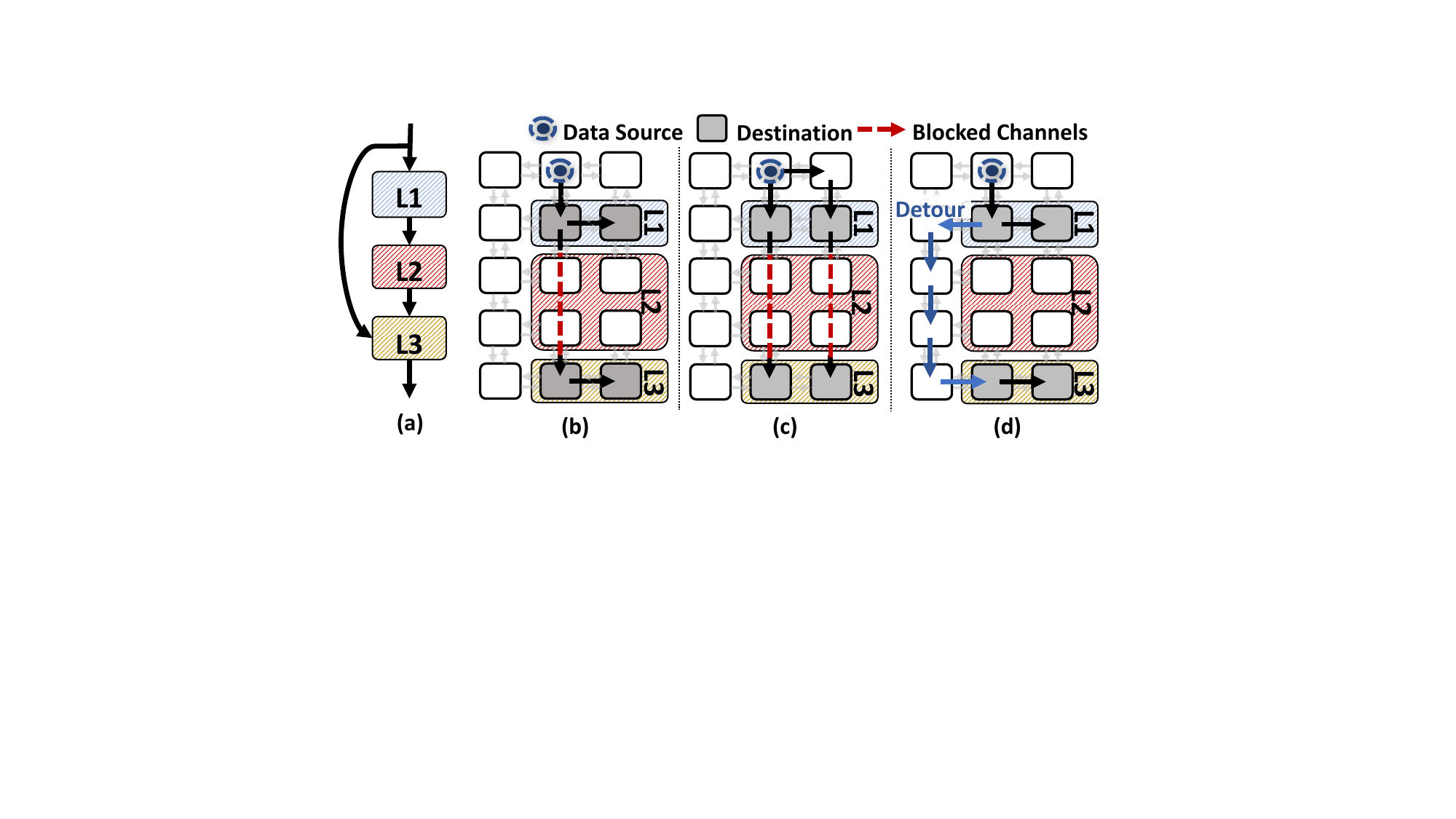}
  \caption{The hardware routed paths for residual blocks. (a) is the residual block to route. (b) and (c) are possible paths routed by RPM~\cite{rpm} and bLDBDR~\cite{bldbdr} respectively. (d) is the path that detours the blocked channels. }
  \label{fig:hardware-limit}
\end{figure}
A straightforward solution is to reserve sufficient hardware resources to avoid potential congestion. For example, the reconfigurable network implements
over-connected hardware with redundant routers and links~\cite{maeri}. Given a task, the network is reconfigured by tailoring these redundant connection resources according to the traffic patterns.  Although it achieves sufficient bandwidth, the tailored network is severely under-utilized and poorly scalable. The reasons have two folds.

First, to enable reconfigurability, the network employs a lot of high-radix routers and wide channels. These costly designs introduce significant area overhead and implementation complexity. In addition, the overhead scales linearly with the network size. 
Second, tailoring the network wastes a great portion of these costly hardware components. Only parts of the links are activated. 

Many approaches have been proposed to improve the efficiency of transmitting collective messages on the NoC. They normally modify routers and routing algorithms to fork and merge flits within the NoC. They provide two primary message patterns, i.e. multicast and reduce, and build other collective communication patterns based on them. For multicast patterns, the tree-based approaches build a virtual multicast routing tree~\cite{carpool, bam, rpm} connecting multiple destinations and fork flits at the branching routers. For reduce traffic patterns, this process is reversed, and some of them combine the flits to reduce within routers~\cite{whirl, carpool}.

The tree-based approaches can alleviate redundant flits of multicast and reduce traffic patterns. However, their traffic scheduling (i.e. {\bf path routing and flow control}) strategies have intrinsic limitations. First, due to the area and timing constraints, a hardware-based traffic scheduling strategy is normally simple and rigid. Second, a router is ``nearsighted'' and works independently. It means that a router may only sense the traffic status around it and make the ``best choice'' by itself. Thus, the hardware cannot co-operate with other routers to optimize its traffic scheduling strategy according to the global status.  

An example in Figure~\ref{fig:hardware-limit} shows the limitation of hardware-based traffic scheduling. Three tensor operators, L1, L2, and L3, form a residual block. The region that L2 occupies easily becomes the hotspot since it is the common space to deliver the output of L1 and the input of L3. 
The input from the data source is fed to both L1 and L3, as shown in Figure~\ref{fig:hardware-limit}~(a). 
With typical hardware-based approaches, the message from the source to L3 should go through the L2 occupied region, though the actual passing ways differ. Figure~\ref{fig:hardware-limit}~(b) and~(c) illustrate two ways for the message to go through the region, and both of them are blocked due to the hotspot. Figure~\ref{fig:hardware-limit}~(d) illustrates a straightforward way to detour the hotspot region. However, the hardware scheduling approaches are hard to adopt this routing way since the detouring builds a non-minimal path.

\subsection{Opportunities for Software Traffic Scheduling}
The principle of solving the aforementioned problem is straightforward. We should adjust the traffic scheduling strategies to mitigate congestion and improve NoC utilization. Though the principle is simple, it is impractical for hardware routers to obtain the run-time traffic information, such as the traffic causing congestion and the under utilized paths around the congestion area. The reason is two-folded. First, the location and intensity of congestion changes temporally due to the short-term bursty patterns discussed in subsection~\ref{subsec:traffic}. Second, different tensor applications may be paralleled and mapped to different cores. It means that the congestion also changes spatially. 


The challenge for hardware can be easily overcome, when we seek a solution at the software level. The rational is that the software has already been responsible for the tensor application's paralleling and mapping to the spatial accelerator. Thus, the software can also obtain the communication information among cores. In other words, we can count on the software to analyze the communication information in advance and find proper traffic scheduling strategies globally across the spatial accelerator.





\section{Software Framework}
Given a tensor application, the goal of our software framework is to find a proper traffic scheduling strategy on the spatial accelerator. To achieve this goal, we first construct a design space for all valid traffic scheduling strategies. Then, we propose a design space exploration method to find the proper solution. In addition, we propose a method to control the traffic injection time at the software level to avoid blocking time caused by congestion.

\subsection{Design Space Formulation}   \label{sec:formulation}
We first formulate three prerequisites, the tensor application, the NoC design, and mapping of the tensor application on the spatial accelerator, in mathematical representations, respectively. Having these representations, we can further formulate path routing and flow control to construct the design space of traffic scheduling strategies.





\paragraph{Tensor application.} A tensor application is represented by a directed hypergraph $G = \left \langle V, E \right \rangle$. $V = \{v_1, v_2, ..., v_n\}$ is the vertex set and $E = \{e_1, e_2, ..., e_m\}$ is the edge set. Each vertex $v_i \in V$ denotes a tensor operator. Each $e_i \in E$ is a \textit{directed hyperedge}, which represents a message to transfer intermediate tensors among operators.  
In the following discussion, the source and destination vertices of a hyperedge $e_i$ are defined as $S(e_i)$ and $D(e_i)$, respectively. 



\paragraph{NoC design.} 
A NoC design is represented by a strongly connected directed graph, $I = \left \langle R, L \right \rangle$. Here, $R = \{r_1, r_2, ..., r_p\}$ is the vertex set and $L = \{l_1, l_2, ..., l_q\}$ is the edge set. Each vertex $r_j \in R$ represents a router, which is bounded to a processing core. Each directed edge $l_j \in L$ is a link connecting two routers in $R$. 



\paragraph{Mapping a tensor application.} The mapping function determines how to allocate each operator to a core of the spatial accelerator. Since a core is bonded to a dedicated router, it is equivalent to mapping each operator to a router mathematically. Thus, given the operator set $V$ and the router set $R$, the mapping function is defined as $f: V \rightharpoonup R$. 



\paragraph{Path Routing.} The path routing determines the path for every message. 
We use $\pi_e$ to represent the valid path for a message $e \in E$. It satisfies the following constraints.  
\begin{enumerate}[label=(\roman*)]
    \item $\pi_e = \left \langle R_{\pi_e}, L_{\pi_e} \right \rangle $ is the sub-graph of $I$.
    \item $\forall s \in f(S(e)),\ \forall d \in f(D(e))$, $\exists$ path $p \subseteq \pi_e$ from $s$ to $d$.
    \item $zi(\pi_e) \subseteq f(S(e)) \wedge zo(\pi_e) \subseteq f(D(e))$
\end{enumerate}
\label{def:path}
\noindent Here, $zi$ and $zo$ represent the nodes with zero input degree and output degree, respectively. 
Constraint \textit{(ii)} indicates that the path should connect every source of the message to every destination.
Constraint \textit{(iii)} guarantees the minimum of the path. All nodes with no input edges should be the message sources. All nodes with no output edges should be the message destinations. 
We further formulate the path routing strategy as follows.
Given a tensor application $G = \left \langle V, E \right \rangle$ and a NoC design $I = \left \langle R, L \right \rangle$, a routing strategy $\Pi_{G, I} =\left \{ \pi_e | e \in E \right \}$ contains a valid path for every message $e \in E$.



\paragraph{Flow control.}
The flow control resolves traffic conflicts when multiple packets at a router request for the same channel. We formulate this problem as a set permutation problem by adopting the message priority mechanism. Every message $e \in E$ is assigned with a unique passive integer $p_e$ as the priority. When two messages $e_i$ and $e_j$ ($p_{e_i} > p_{e_j}$) try to pass same channel simultaneously, the channel conflict happens. The message $e_i$ with higher priority passes successfully, while the other message $e_j$ is blocked to wait until all flits of $e_i$ pass through. 
The flow control is a permutation $P_{|E|}$ of the integer set $\left \{1, 2, ..., |E| \right \}$, where the $i$-th factor dictates the priority of the message $e_i$. 






\paragraph{Space Representation.}   \label{subsubsec:represent}
We need a representation to simultaneously embeds the space of path routing and flow control. Moreover, the representation should not diminish the design space of any of them. Fortunately, the formulations proposed are orthogonal to each other: changing the routing has no effect on the flow control strategies, and vise-versa. Consequently, a straightforward way to build the representation is to concatenate them together.
Given a tensor application $G = \left \langle V, E \right \rangle$ and an interconnection network $I = \left \langle R, L \right \rangle$, a traffic scheduling strategy is a pair of $Stg = (\Pi_{G, I}, P_{|E|})$.

\subsection{Strategy Sampling Technique} \label{subsubsec:sample}
Before design space exploration, we need to solve the problem of how to sample a new traffic scheduling strategy $Stg = (\Pi_{G, I}, P_{|E|})$ from the design space efficiently. The problem is challenging because the routing strategies are hard to get from the scratch: sub-graphs of $I$ that meet all constraints in Def.~\ref{def:path} are just a small portion of the entire space. Thus, randomly sampling the routing path from the entire space is very inefficient. 

\begin{algorithm}[h]
  \SetAlgoLined
  \KwData{A valid path $\pi_e = \left \langle R_{\pi_e}, L_{\pi_e} \right \rangle$ of message $e$}
  \KwResult{A new valid path $\pi_e' = \left \langle R_{\pi_e'}', L_{\pi_e'}' \right \rangle$ of $e$}
    \tcp{remove a weakly connected sub-graph}
    $IM_{\pi_e} \leftarrow \pi_e - f(S(e)) - f(D(e))$\;
    $I = \left \langle R_I, L_I \right \rangle \subseteq_R IM_{\pi_e}$ that $I$ is weakly connected\;
    $\pi_e' \leftarrow \pi_e - I$\;
    \tcp{add a new node $r'$ out of the path}
    $r' \in_R \overline{R_{\pi_e}}$\;
    $R_{\pi_e'}' \leftarrow R_{\pi_e'}' + {r}$\;
    \tcp{dangling nodes}
    $D^+ \leftarrow \{u | uv \in L_{\pi_e} \wedge v \in R_I \wedge u \notin R_I \}$, $D^- \leftarrow \{v | uv \in L_{\pi_e} \wedge u \in R_I \wedge v \notin R_I \}$\;  
    $\pi_e' \leftarrow \pi_e' \cup \{ XY\ connects\ from\ d^+\ to\ r'\ |\ d^+ \in D^+\} \cup \{ XY\ connects\ from\ r'\ to\ d^-\ |\ d^- \in D^-\}$\;
    check if $\pi_e' \subseteq I$\;
    remove circles from $\pi_e'$\;
    
  \caption{Rebuild a routed path}
  \label{algori:route}
\end{algorithm}

To solve this problem, we generate a new scheduling strategy by disturbing a valid strategy randomly. Given an original valid strategy $Stg = (\Pi_{G, I}, P_{|E|})$, we propose two disturbing technique to generate a new $P_{|E|}$ or a new $\Pi_{G, I}$, respectively. 
{\bf Flow control disturbing} generates a new $P_{|E|}$ by reordering the priorities of two randomly selected messages in $P_{|E|}$ .
{\bf Path routing disturbing} generates a new $\Pi_{G, I}$. It randomly selects a path $\pi_e \in \Pi_{G, I}$ and rebuilds an intermediate sub-path of it. 

Algorithm~\ref{algori:route} illustrates the details. When rebuilding the routed path $\pi_e$, we first randomly remove a weakly connected sub-path $I \subseteq \pi_e$ that contains neither sources nor destinations (\textit{line 1-3}). 
Then, we randomly select another router $r'$ from the complement node set $\overline{R_{\pi_e}}$ (\textit{line 4-5}) and connect it to the path $\pi_e$. To do so, we add XY connects between $r'$ and all the dangling nodes (\textit{line 6-7}).
Finally, we detect and eliminate the circle of the resulting path $\pi_e'$ (\textit{line 8}).

\subsection{Design Space Exploration}
In order to pursue an optimal traffic scheduling of a whole tensor application, we propose a heuristic searching algorithm based on the Evolutionary Algorithm (EA). It employs the formulations in \ref{subsubsec:represent} to encode the strategies, and uses the space sampling technique in \ref{subsubsec:sample} to mutate individuals. It takes the overall execution cycles to evaluate the fitness. 
We describe the details of the EA algorithm in the following.

\noindent \textbf{Initialization.} We initialize the algorithm by mutating several valid scheduling strategies from existing works~\cite{bam, whirl, rpm, steiner}. Specifically, to initialize an individual, we generate a flow co    ntrol strategy by randomly permuting the $P_{|E|}$. Then, we build the routing strategy in two steps. First, we randomly select a routing strategy from BAM~\cite{bam}, whirl~\cite{whirl}, RPM~\cite{rpm}, and Steiner minimal tree algorithm~\cite{steiner}. Second, we disturb it by randomly picking $k$ routes and applying {\bf path routing disturbing} on them. 

\noindent \textbf{Mutation - path routing strategy.} Given a scheduling strategy $Stg = (\Pi_{G, I}, P_{|E|})$, we mutate the path routing strategies by randomly picking $k$ routes from $\Pi_{G, I}$ and applying the \textit{path routing disturbing} on each of them. The parameter $k$ is set by the mutation rate of the algorithm.

\noindent \textbf{Mutation - flow control strategy.} Given a scheduling strategy $Stg = (\Pi_{G, I}, P_{|E|})$, we mutate flow control strategies by randomly picking $k$ pairs of $P_{|E|}$ and applying the \textit{flow control disturbing} on each pair, with the same parameter $k$ of path routing strategy mutation.

\noindent \textbf{Crossover.} Crossover produces a new solution by inheriting the well-performed characteristics of parents. We randomly select two solutions from parent population subsets and blend their characteristics in two phases. First, we interchange flow control strategies between parents. Second, we randomly select a half of messages and interchange their corresponding paths in parents' path routing strategies. 

\noindent \textbf{Fitness Evaluation. } After crossover and mutation, an evaluation process measures the population's fitness. We evaluate the fitness of a traffic strategy considering two factors: the overall application performance and the bits for storing controlling information. To get overall performance, traditional approaches include a cycle-accurate simulation or a statistical analysis model.
However, these approaches are either time-consuming or suffer low accuracy. In this work, we leverage the deterministic features of tensor applications. 
We build a performance model to calculate the execution time with high accuracy, which is further discussed in the section~\ref{subsec:block_time_shift}.


\subsection{Injection Time Control}    \label{subsec:block_time_shift}
The tree saturation problem is one of the major causes of network performance degradation. When a packet is blocked in the network, its injected flits wait on routers and take up the channels. In turn, these flits further prevent other packets to pass the channels they hold, causing head-of-line blocking. Due to the queuing effect, the head-of-line blocking may spread to channels and routers nearby, forming a tree-like saturation region. 


To mitigate this problem, we propose the \textit{Injection Time Control} technique at the software level. The goal is to delay injecting the messages, which will be blocked during traversing. Thus, the messages wait in the computing core instead of in the network. We can release the channels, which are held by the originally blocked message. Hence, we prevent the head-of-line blocking problem. 

The algorithm is simple, we maintain unavailable cycles for every channels. Given a message $e$ to issue, we check the its path $L_{\pi_e}$ to see whether its passing channels are held by other messages when $e$ tries to pass through. If so, we delay the injection time of $e$, and let its source core temporally buffer the message until all channels of the whole path are free.


\begin{figure}[htb]
  \centering
  \includegraphics[width=0.5\textwidth]{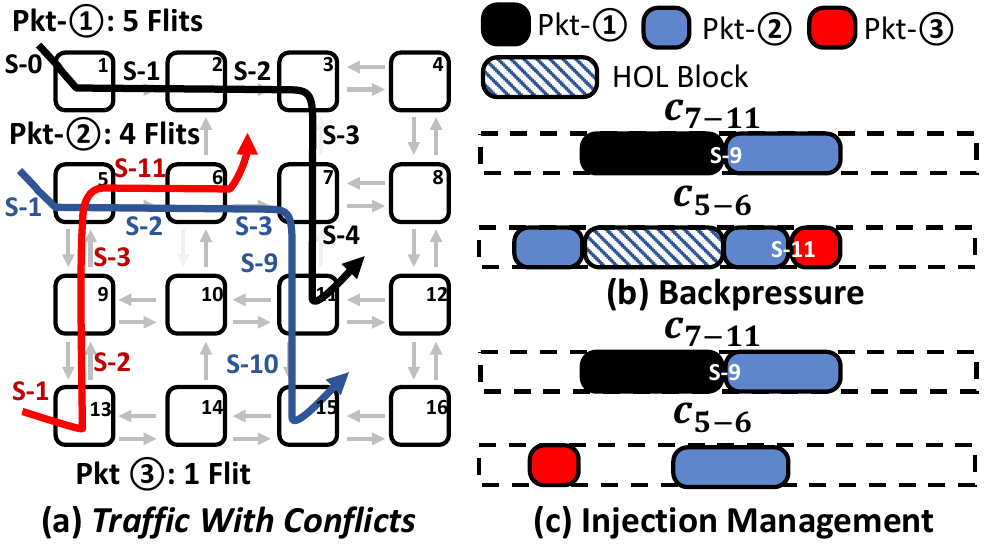}
  \caption{Scheduling Methods to Solve Conflicts.}
  \label{fig:pp-blk-shift}
\end{figure}

We use the example in Figure~\ref{fig:pp-blk-shift} to illustrate this technique. For simplicity, we use the ``time slot'' to represent the time, when a flit traverses over a hop. 
For example, S-1  in the figure denotes time slot-1. In this example, there are three packets (\textcircled{1}, \textcircled{2}, \textcircled{3}) to be injected at slot-0, slot-1, and slot-1, respectively. We assume their priorities with the order of $P\left [\textcircled{1} \right ] > P\left [\textcircled{3} \right] > P\left [\textcircled{2} \right]$. As shown in Figure~\ref{fig:pp-blk-shift}~(a), packet-\textcircled{1} and packet-\textcircled{2} conflict at channel 7-11 ($c_{7-11}$). Packet-\textcircled{2} and packet-\textcircled{3} conflict at channel 5-6 ($c_{5-6}$).

We first consider a baseline back-pressure flow control. Once blocking happens, a router informs its adjacent upstream nodes to stop them from transmitting flits. Initially, every channel is free at slot-0. As shown in Figure~\ref{fig:pp-blk-shift}~(a), packet-\textcircled{1} (the black one) is injected at slot-0, then passes the channel $c_{7-11}$ at slot-4. However, at the same slot, the packet-\textcircled{2} (the blue one) also tries to pass the channel $c_{7-11}$. According to the priority order, packet-\textcircled{2} wins the conflict and holds $c_{7-11}$ until its last flit passes through at slot-8. On the contrary, packet-\textcircled{2} is blocked in the network. Hence, the already injected flits of packet-\textcircled{2} occupy the channel $c_{5-6}$ from slot-2 to slot-10 because of the head-of-line blocking, as illustrated in Fig.~\ref{fig:pp-blk-shift}~(b). During this period, packet-\textcircled{3} tries to pass the channel $c_{5-6}$ at slot-3. But it finds that the channel has already been held by packet-\textcircled{2}. Consequently, the packet-\textcircled{3} waits in the network due to the head-of-line blocking caused by $c_{7-11}$. In summary, it takes packet-\textcircled{3} 12 slots to traverse over the network. 

Then we consider the \textit{Injection Time Control} technique. As shown in Fig.~\ref{fig:pp-blk-shift}~(c), since we predict that packet-\textcircled{2} will be blocked to slot-8 according to the priority order, we should not inject this message into the network as early as slot-1. Instead, we delay injecting it at slot-6 to let its first flit pass channel $c_{7-11}$ at slot-9, as soon as packet-\textcircled{1} releases this channel. With such a method, the packet-\textcircled{2} does not hold the channel $c_{5-6}$ until slot-7. Consequently, packet-\textcircled{3} can traverse over $c_{5-6}$ at slot-4 successfully. In summary, with the \textit{Injection Time Control} technique, we reduce the delay of packet-\textcircled{3} from 12 slots to 5 slots, without inducing extra delay for other packets.

Since the \textit{Injection Time Control} technique determines the messages' injection time and guarantees the messages' receiving time, we can build an accurate analysis model based on it. The model employs a tensor analysis framework, such as Timeloop~\cite{timeloop}, to calculate the computation cycles for tensor operators, and leverages \textit{Injection Time Control} to determine communication cycles for cross-operator messages. This model helps the scheduling space exploration with an accurate performance evaluation.

\section{Hardware Design} ~\label{sec:hardware}
The router design should be modified to work with the software traffic scheduling strategy. On the one hand, the original components designed for hardware traffic scheduling can be simplified or even removed. On the other hand, the software traffic scheduling strategy may consume hundreds or even thousands of bits of information. A dedicated architecture is required to reduce the overhead of applying the strategy. 


In this section, we first introduce the pipeline architecture of the router and compare it with a traditional design to demonstrate the difference. Then, we propose a \textbf{hybrid routing} mechanism to reduce the traffic scheduling information carried in the packet headers. In addition, we adopt the \textbf{trunk-level message} to mitigate the redundant packet headers.




\subsection{Router Pipeline}

\begin{figure}[h]
  \centering
  \includegraphics[width=0.48\textwidth]{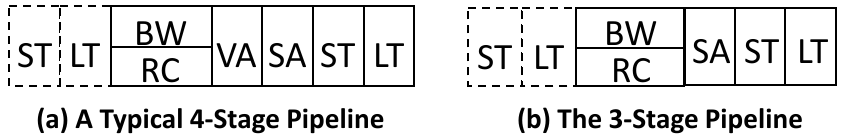}
  \caption{The router pipeline}
  \label{fig:router-pipe}
\end{figure}

A typical four-stage pipeline for the traditional multi-channel router architecture is shown in Fig.~\ref{fig:router-pipe}~(a). Once a flit reaches the router after traversing through the link (LT), it is written to a buffer (BW) and performs routing computation (RC) at the same time in the first stage. Then, it raises a request for granting virtual channels at the next hop (VA) in the second stage. After that, it requests for allocating the switch (SA) in the third stage. At last, it traverses through the switch (ST) and goes through the output link (LT) to the next router. 

The pipeline stage of our router design is shown in Fig.~\ref{fig:router-pipe}~(b). The first stage is also responsible for buffer write (BW) and routing computation (RC). However, as the routing strategy has been given by the software. The hardware for this part is re-designed to support a hybrid routing mechanism. It can apply the software routing strategy efficiently while reducing the information carried in each packet. More details are provided in section~\ref{subsec:hybrid_routing}. 

The original virtual channel allocation (VA) stage is removed from the pipeline. The reason is explained as follows. Virtual channels are introduced to resolve the head-of-line problem mentioned in section~\ref{subsec:block_time_shift}. However, this problem has already been resolved at the software level with the help of \textit{Injection Time Control} technique. Thus, we can remove the virtual channels and the corresponding allocators and arbitrators, which reduce hardware design complexity significantly.

\subsection{Hybrid Routing} \label{subsec:hybrid_routing}
 
\begin{figure}[htb]
  \centering
  \includegraphics[width=0.48\textwidth]{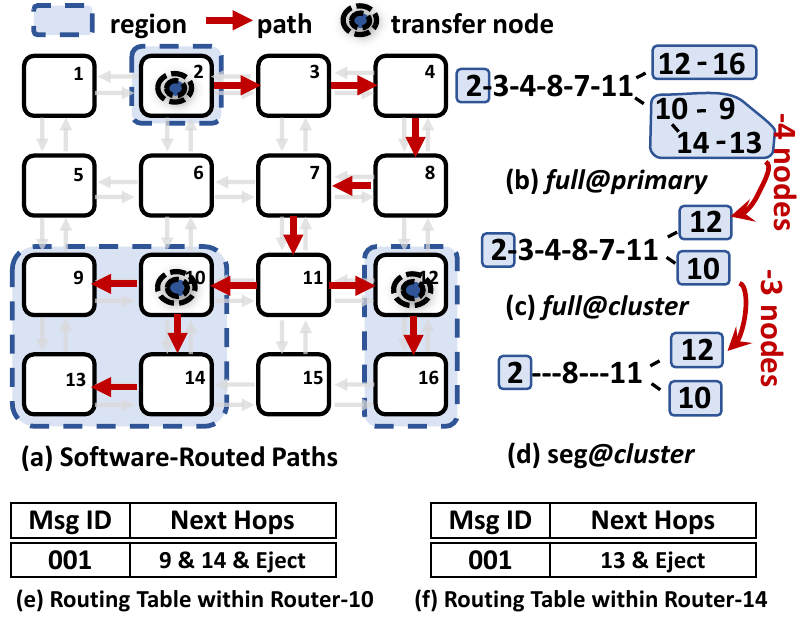}
  \caption{An example of the hybrid routing mechanism.}
  \label{fig:routing_methods}
\end{figure}

In this section, we describe a hybrid routing mechanism to compress the routing information. We use the routing path in Figure~\ref{fig:routing_methods}~(a) as an example to illustrate this compression method. In this example, a packet is sent from the source (router 2) to multiple destinations (router 10,9,14,13,12,16). We start from a straightforward source routing mechanism that stores all routing information in the packet header, as shown in Fig.~\ref{fig:routing_methods}~(b). We term such recording mechanism as \textit{full@primary}. \textit{full@primary} merges the common path that all the flits from every source to every destination would have traversed. \textit{full@primary} is poorly scalable because it involves linear complexity with respect to the network size. The long packet header wastes the network bandwidth and enlarges the packet traversal latency. In the following discussions, we propose two methods to reduce the routing information within packet headers.




\begin{figure}[b]
  \centering
  \includegraphics[width=0.48\textwidth]{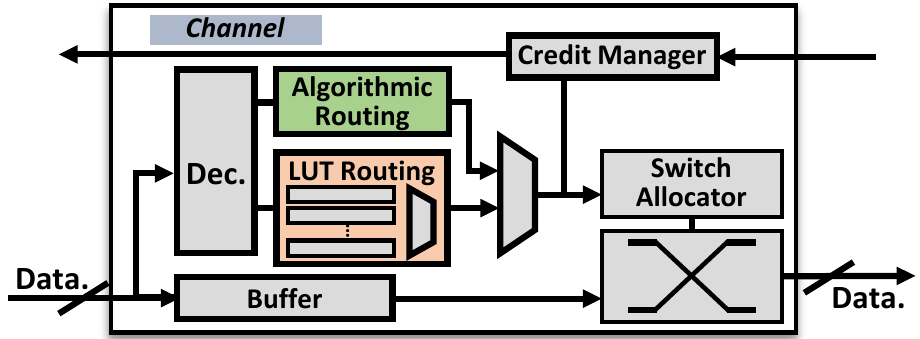}
  \caption{Router micro-architecture. The algorithmic routing module performs dimension-order routing. The LUT routing module performs table-based routing.}
  \label{fig:router-overview}
\end{figure}

\paragraph{Node-table routing.}
We first eliminate to encode the entire source destination list within the header. As shown in Figure~\ref{fig:routing_methods}~(a), we cluster the adjacent sources and destinations together as a \textit{region}, and select a node in the region as the hub node. In this case, node-2, node-10, and node-12 are selected as the hub nodes of their corresponding regions, respectively. A header only records the routing information between hub nodes. As Figure~\ref{fig:routing_methods} illustrates, we remove the other nodes within the region from the header routing information. Instead, tabled-based routing determines the paths of packets entering regions. To achieve this, we add a routing table at each router with pre-programmed routing entries (Figure~\ref{fig:router-overview}). Figure~\ref{fig:routing_methods}~(e) and~(f) illustrate the entries of routing tables, which record the message's next hops. In this case, when the packet reaches node-10, the header has no further routing information to choose the next hop. Node-10 thus checks its routing table with the message id and forwards the packet to node-9 and node-14 and the core it attaches from the network. We term this method as \textit{full@cluster}, which prevents the header to store the routes within regions. In this case, we reduce the length of source routing by four nodes.  




An important design parameter of \textit{full@cluster} is the size of a routing table. For a node's routing table, its number of routing entries is equal to the total number of messages that are sent by or sent to that node. This parameter is determined by the type and the number of tensor operators mapped to each core. We can take the deep neural network as an example. For such a tensor application, each operator takes two tensor as the input and one tensor as the output, which result in three messages on each node. Thus, the size of routing table should be larger than $3\times N_{op}$. Here, $N_{op}$ denotes the maximum number of operators assigned to a core. In fact, for large tensor applications, it is really common that only one operator is assigned to each core. Thus, the overhead of this routing table is trivial. In addition, the software can limit the number of operators mapped to each core so that the total messages can be recorded in the routing table. 


\paragraph{Algorithmic routing.}
To further compress the routing information, instead of storing every hops along the routes, we only record multiple ``critical nodes'' of them. Specifically, we divide the route into multiple segments; the cut points of these segments are critical nodes. Other hops along the segments are inferred with algorithmic routing based on the critical nodes. For the path 2-3-4-8-7-11 in Figure~\ref{fig:routing_methods}~(d), we maintain node-2, node-8, node-11 as the ``critical nodes'' and store them within the packet header. The path is thus divided into two segments: the segment from node-2 to node-8, and the segment from node-8 to node-11. For the segment from node-2 to node-8, the intermediate routes, e.g. node-3, are calculated by dimension-order routing (XY routing in 2D mesh) with the endpoints of node-2 and node-8. We term this method as \textit{seg@cluster}. Compared with \textit{full@cluster}, \textit{seg@cluster} prevent to encode every hops along the path, which reduces the length of source routing by three nodes.


\paragraph{Architectural supports.}
To enable this hybrid routing mechanism, we integrate two routing modules within the router, as Figure~\ref{fig:router-overview} illustrates. The algorithmic routing module calculates the routes based on the simple logic (dimension-order routing in this work), and the LUT routing module performs table-based routing. Accordingly, the router works in two modes: the pop-then-calculate mode and the look-up-table mode. In the pop-then-calculate mode, a routing module extracts the first ``critical node'' from the header, then calculates the output channels based on its relative location to the critical node. Once a critical node is reached, it is popped out from the path. Once the source routing information is empty, the router changes from pop-then-calculate to the look-up-table mode. In the look-up-table mode, a routing module simply checks its routing table with the message id and then gets the corresponding output channels.

\begin{figure}[h]
  \centering
  \includegraphics[width=0.48\textwidth]{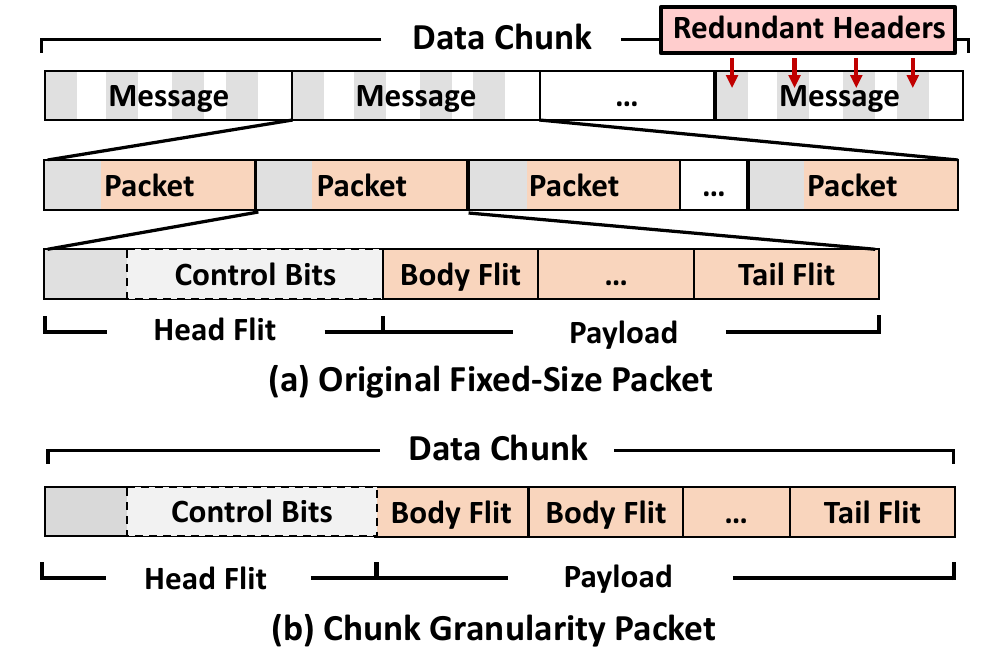}
  \caption{Packetization Schemes.
  (a) Massive small fixed size packets to carry a data chunk and (b) a single packet with the size of the entire data chunk. }
  \label{fig:pp-packet}
\end{figure}

\subsection{Chunk Granularity Packet} \label{subsec:packet}

In typical interconnection networks, as shown in Figure~\ref{fig:pp-packet}~(a), a large data trunk is first divided into small messages, then a message is further split into multiple fixed-size packets. A packet is the basic traffic granularity that stores the scheduling information within its head flit and carries data with its body flits. For \name, such a packaging scheme is inefficient in two folds: 
First, the messages for the same data chunk are scheduled independently by the framework. Such an approach not only increases the search space, but also introduces multiple head flits to control every single message. 
Second, the packets within a message share the same scheduling information, but each of them requires an individual head flit. The redundant head flits waste the bandwidth. 


To reduce such overhead, we extend the message-based forwarding idea in~\cite{communication-isca21} and increase the packet scale to the entire data chunk. We collapse the hierarchy of chunk, message, and packet, resulting in a single packet to deliver the whole data chunk, as Figure~\ref{fig:pp-packet}~(b) shows. We use a single head flit to signify the start of the data chunk transmission, which sets up the passing channel states. And we explicitly introduce a tail flit to release the channels after transmission. This chunk granularity packet can reduce the overhead of headers, thereby improving the bandwidth efficiency.

\section{Evaluation}    \label{sec:evaluation}


\subsection{Experimental Setup}
We implement the proposed \name's software scheduling framework using Python and C++. The inputs of the framework include the mapped tensor application operator graph and the NoC specifications. The framework automatically performs task scheduling and generates low-level hardware control configurations. 
To support the hardware proposed in Section~\ref{sec:hardware}, we modify the cycle-accurate Booksim simulator~\cite{booksim} to model the router pipelines and micro-architectures. 




\begin{table}[h]
\centering

\resizebox{0.47\textwidth}{!}{%
\begin{tabular}{c|c|c|c}
\hline
\textbf{Name} & \textbf{Tasks} & \textbf{\begin{tabular}[c]{@{}c@{}}Pipelined \\ Layers\end{tabular}} & \textbf{\begin{tabular}[c]{@{}c@{}}Tensor Operators \\ Per Layer\end{tabular}} \\ \hline \hline

\multirow{4}{*}{\textbf{Multi-Model-1}} & ssd   & 16 & 8  \\ \cline{2-4} 
                                            & resnext50-32x4d & 16 & 8  \\ \cline{2-4} 
                                            & mnasnet        & 16 & 8  \\ \cline{2-4} 
                                            & bert-large           & 8  & 8  \\ \hline \hline
\multirow{4}{*}{\textbf{Multi-Model-2}} & bert large   & 8 & 8  \\ \cline{2-4} 
                                            & unet & 12 & 8  \\ \cline{2-4} 
                                            & inception        & 24 & 8  \\ \cline{2-4} 
                                            & wide-resnet50           & 24  & 8  \\ \hline \hline
\multirow{4}{*}{\textbf{Multi-Model-3}} & wide-resnet50   & 16 & 8  \\ \cline{2-4} 
                                            & resnext50-32x4d & 16 & 8  \\ \cline{2-4} 
                                            & resnet50        & 16 & 8  \\ \cline{2-4} 
                                            & vgg16           & 4  & 8  \\ \hline \hline
\textbf{Mobilenet-v3}                      & mobilenet\_v3   & 64 & 8 \\ \hline \hline
\end{tabular}%
}
\caption{Tested Benchmarks}
\label{tab:workload}
\end{table}

We evaluate \name~on the most popular tensor tasks, namely deep neural networks. To make full use of  the  large-scale spatial accelerators' computation power  and fully exhibit \name's performance, we consider the challenging multi-model serving tasks~\cite{multi_nn,ghodrati2020planaria}. Multiple  DNN models run in parallel. Multiple layers within a model run in a pipelined manner. As listed in Table~\ref{tab:workload}, we select nine DNN models~\cite{bert, wide_resnet, resnet, resnext, vgg, unet, mnasnet, inception, ssd} from multi-model applications, VR/AR tasks, and MLPerf~\cite{mlperf}. We adopt the implementations in torchvision 0.9.1 for WideResnet, Resnet, ResneXt, Vgg, Unet, Mnasnet, Inception and SSD. We use Bert and Bert-large from Hugging Face’s Transformers library~\cite{transformer_lib}. To fit into the spatial architecture, we divide the models into multiple fixed-size segments. Layers within a segment are processed pipelined, and segments of a model are processed one by one. We allocate a fixed group of cores for each model, which does not change during the execution.


We develop a compiling toolchain to automatically generate the tensor operator graph from high-level tensor application descriptions, e.g. the ONNX format DNN models. The toolchain employs Timeloop~\cite{timeloop} to map a tensor operator to the cores. It partitions the core array based on the Hilbert Curve and allocates the partitions to these operators. Once a core is mapped with multiple operators, it executes them in an out-of-order style that issues the operator once all its operands are received.

As shown in Table~\ref{tab:hw_setup}, the spatial accelerator consists of a $16 \times 16$ core array. Each core contains a $16 \times 16$ MAC array. Each MAC performs one 8-bit multiply-addition per cycle and integrates 56 bytes register to temporally hold operands and partial sums. This MAC array is organized in an NVDLA-like weight stationary dataflow~\cite{simba}. Each core is also equipped with a 260 KB SRAM scratched buffer, which is private and could only be accessed by components within the core. 



Both the NoC and computation tiles work at 500 MHz. And we adopt the same High Bandwidth Memory (HBM) employed by TPUv4~\cite{tpuv4} as the main memory, which provides up to 1200 GB/s memory bandwidth. We assume that the system has 8 memory controllers, which equally divide the bandwidth. These memory controllers are connected to the cores at the middle of four edges. For both baselines and our design, we set the array topology as a 2D mesh. Note that our work also supports arbitrary interconnection typologies. 

\begin{table}[t]
\centering

\resizebox{0.42\textwidth}{!}{%
\begin{tabular}{cc|c}
\hline 
\multicolumn{2}{c|}{\textbf{Parameter}}                                & \textbf{Configuration} \\ \hline \hline 
\multicolumn{1}{c|}{\multirow{2}{*}{MAC}}     & Operation Precision     & 8 bits                 \\ \cline{2-3} 
\multicolumn{1}{c|}{}                        & Register           & 56 Bytes                  \\ \hline \hline 
\multicolumn{1}{c|}{\multirow{4}{*}{Core}}   & MAC Array               & $16 \times 16$                \\ \cline{2-3} 
\multicolumn{1}{c|}{}                        & Private Buffer          & 260 KiB                \\ \cline{2-3} 
\multicolumn{1}{c|}{} & Dataflow & \begin{tabular}[c]{@{}c@{}}NVDLA-Like\end{tabular} \\ \cline{2-3}
\multicolumn{1}{c|}{}                        & Core Clock                   & 500 MHz                  \\ \hline \hline 
\multicolumn{1}{c|}{\multirow{3}{*}{System}} & Core Array              & $16 \times 16$               \\ \cline{2-3} 
\multicolumn{1}{c|}{}                        & HBM Bandwidth           & 1200 GBps               \\ \cline{2-3} 
\multicolumn{1}{c|}{}                        & Mapped Controllers & 8                      \\ \hline \hline
\multicolumn{1}{c|}{\multirow{3}{*}{Network}} & Topology              & Mesh           \\ \cline{2-3} 
\multicolumn{1}{c|}{}                        & Router Clock         & 500 MHz              \\ \cline{2-3} 
\multicolumn{1}{c|}{}                        & Flit Width & 1024 bits \\ \cline{2-3} \hline \hline 
\end{tabular}%
}
\caption{Hardware Setups}
\label{tab:hw_setup}
\end{table}


We use Timeloop~\cite{timeloop} framework to estimate the latency and energy of computing cores, with the back-end of gem5-Alladin~\cite{gem5} and CACTI~\cite{cacti} to model logical circuits and SRAM buffers, respectively.  To evaluate the interconnection network, we extend BookSim2~\cite{booksim}, a cycle-accurate interconnection simulator, to integrate with Timeloop for full-system evaluation. 



\begin{figure*}[t]
  \centering
  \includegraphics[width=\textwidth]{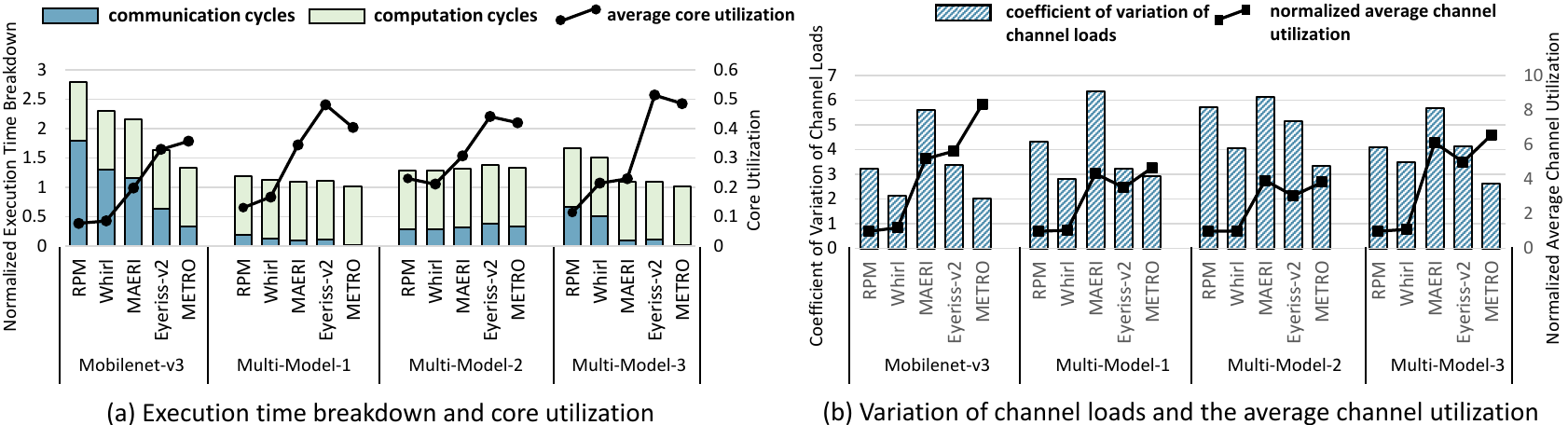}
  \caption{Processing tensor applications on a spatial architecture with $16 \times 16$ computing cores: 
  (a) computation and communication breakdown normalized to the computation time (primary), and average utilization of core's computing time (secondary),
  (b) coefficient of variation of the channel loads (primary) and average channel utilization normalized to the RPM scheduling approach (secondary).}
  \label{fig:performance}
\end{figure*}

We compare \name~with both traditional NoCs and reconfigurable interconnections. 
For traditional NoCs, we assume the baseline bufferable routers with five bi-direction ports. They perform the credit-based back-pressure virtual-cut flow control. We assume four virtual channels (VCs) per physical channel. One of them acts as the escape VC, while the other three VCs are allocated for packets in a round-robin manner. Each VC takes up 2 flits, with 512 bits per flit. The router performs the tree-based multicast, which replicates the flits at the crossbar on the fly. For each multicast flit, the router requests a buffer for each output port. If the request on any port fails, the flit has to stay at the virtual channel buffer and tries again at the next cycle. Based on this router setup, we take two traditional multicast-enabled NoCs as baselines.

\begin{itemize}
    \item Recursive Partitioning Multicast (\textbf{RPM})~\cite{rpm}: RPM partitions the network into eight quadrant based on the relative location of nodes to the source. It routes the packets with this partitioning scheme and a fixed port priority order.

    \item \textbf{Whirl}~\cite{whirl}: Whirl attempts to balance the traffic load by randomly selecting the branching styles for multicast packets. And it optimizes the router pipeline to at least 1-cycle. Noted that we only consider the 1-to-M network of Whirl, but not include its ack-combination network. It is because the packets of tensor application cannot be reduced with simple combination operations as adopted in Whirl. 

\end{itemize}

\noindent For reconfigurable networks, we employ MAERI~\cite{maeri} and Eyeriss v2~\cite{eyerissv2} as the baselines. For fairly comparison, we scale the nodes that the network connects (a multiplier switch in MAERI or a PE in Eyeriss v2) to a computing core as shown in Table~\ref{tab:hw_setup}. Noted that these reconfigurable networks are designed for restricted tasks. For example, Eyeriss-v2 only supports the row-stationary paralleled tensor operators layer by layer. It is unable to process multiple operators simultaneously. Consequently, they cannot be directly scaled up for spatial architectures. Hence, we slightly modify them to support the flexible tensor applications in our scenario.

\begin{itemize}
    \item \textbf{MAERI}~\cite{maeri}: MAERI organizes the computing cores as a 1D systolic-like vector. MAERI builds a binary chubby tree to distribute operands, and leverages a data reduction tree to reduce partial sums. We set the bandwidth of the lowest level in the chubby tree to 1024-bit, which is same as the link width in Table~\ref{tab:hw_setup}. To connect 512 computing cores, both trees have nine levels. The bandwidth of a link scales up from the lowest level to the upper levels with factors of 1x-1x-1x-1x-2x-2x-2x-2x-2x. The root link matches HBM bandwidth. All links in the reduction tree are 1024 bits.
    
    \item\textbf{Eyeriss v2}~\cite{eyerissv2}: Eyeriss v2 connects the computing core array with a hierarchical mesh. Cores within a cluster are connected to a router cluster with all-to-all network. And the router clusters are connected as a mesh. Each cluster has $4 \times 4$ computing cores, with four high-radix routers. There are $8 \times 8$ clusters in total. We assume only one network with 512 bit width for fairly comparison. To process different operands, we reconfigure the topology within a cluster, which is not originally supported by Eyeriss v2. 
\end{itemize}

\subsection{Evaluation Results}
\vspace{-1em}

\subsubsection{Performance Improvements:}  \label{subsec:performance}

Figure~\ref{fig:performance}~(a) shows the execution time breakdown for running tensor applications on the spatial architecture, and Figure~\ref{fig:performance}~(b) shows the variation of the channel loads. 
As shown in Figure~\ref{fig:performance}~(a), for traditional multicast-enabled NoCs, tensor applications consume a significant amount of time on communication. Convolution-based applications such as Multi-Model-1, Multi-Model-2, and Multi-Model-3 exhibit high computing intensity that perform more computation on a message, thus induce low traffic intensity. 
On the contrary, Mobilenet-v3 employs intensive depth-wise convolutions, which have less compute intensity to reuse data, leading to communication much more dominant. In summary, communication time can take from 11\% to 64\% of overall execution time for traditional multicast-enabled NoCs. For computation insensitive applications such as Multi-Model-1, \name~improves the overall execution performance by up to 14\% to 33\% compared to traditional NoCs. And for traffic-intensive applications, Mobilenet-v3, \name~improves the overall execution performance 52\%. The \name's improvement on core utilization is more obvious. For the traffic-intensive application Mobilenet-v3, \name~improves the average core utilization by over 5x. For the application with lower traffic intensity, \name~can also improve the utilization by over 1.8x. The improvement of \name~over multicast-enabled traditional NoCs comes from two folds. First, the software scheduling framework performs wiser routing and flow controls to detour the hotspot and leverages the under-utilized channels, it increases the network elasticity to hold the bursty traffic. Second, our co-designed hardware has a shorter controlling pipeline, which not only reduces the traffic delay of packets, but also reduces the channel turnaround time. 


Surprisingly, Figure~\ref{fig:performance}~(a) illustrates that interconnections inspired by reconfigurable networks also exhibit communication bounds. It is due to the intrinsic scaling problem of these reconfigurable networks. 
Since these networks integrate various redundant channels to perform reconfiguration, scaling the networks up will result in a considerable overhead to implement these redundant channels. To be fit into a reasonable hardware overhead, their redundant channels should be tailored when the network size is large. Taking MAERI-like networks as the example. Ideally, data distribution tree in MAERI scales the bandwidth up with factor 2x at each layer from the leave links to the root link. However, for the array with 512 cores, such scaling way requires the root link to provide over 65,000 GB/s bandwidth. Hence, considering the hardware limitations, we just double the bandwidth at the first five levels of the tree, while keep the bandwidth unchanged for the other four levels. 

In Figure~\ref{fig:performance}~(b), we use the coefficient of variation as the metric to evaluate the balance of channel loads. It is the ratio of standard deviation channel loads to the average number of channel loads. The higher the value is, the more imbalance the channel load is. Whirl exhibits good balance on most benchmarks, since it randomly selects the valid paths from the candidates. MAERI-like networks exhibit the worst balance of channel loads, since the tree topology has the intrinsic unbalance problem that the channels closed to the root are much more frequently used than the channels closed to leaves. \name~reduces the coefficient of variation up to 42\% compared with traditional NoCs, and reduces up to 63\% compared with reconfigurable networks. Because \name~software explores much a wider range of the routing space, especially for the channels far from the hotspots. On the contrary, whirl only explores the minimum spanning trees, which are restricted within the rectangles formed by sources and destinations. 

Benefited from balance load, \name~achieves the great improvement on channel utilization. As shown in Figure~\ref{fig:performance}~(b), \name~has up to 8x channel utilization improvement compared with traditional NoCs because of the more efficient scheduling strategies. 
\name~also achieves 1.6x higher channel utilization improvement compared with reconfigurable NoCs because the \name~does not rely on the redundant links to provide high performance. Noted that reconfigurable NoCs have a higher channel utilization rate than traditional NoCs because they have fewer traffic congestions and the resulting tree saturation problem.

\subsubsection{Improvement Breakdown:}      \label{subsec:breakdown}

\begin{figure}[h]
  \centering
  \includegraphics[width=0.5\textwidth]{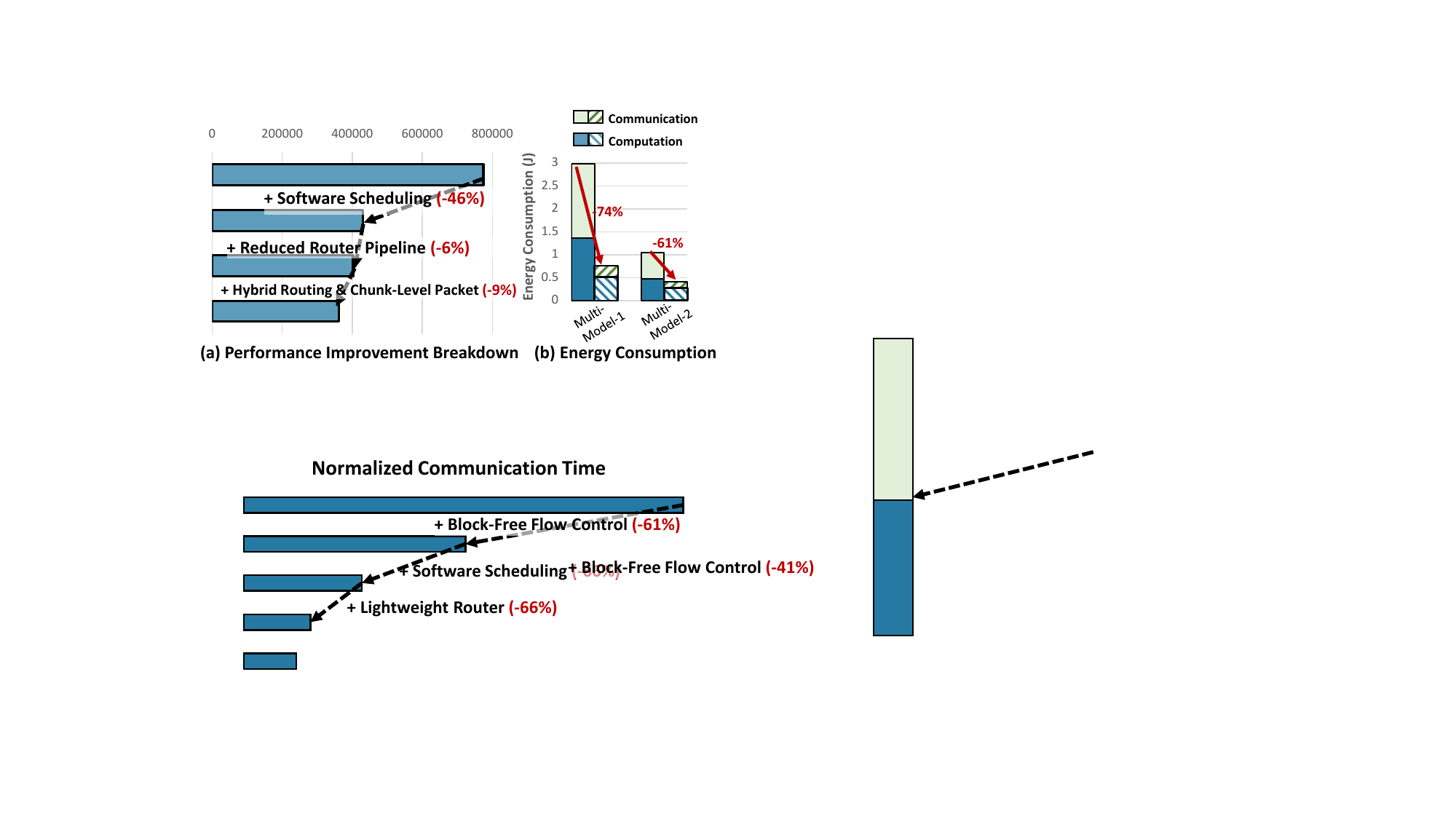}
  \caption{(a) Performance improvement breakdown (b) Energy consumption}
  \label{fig:breakdown}
\end{figure}

To better understand the contributions of different optimization approaches of \name, we separately apply the proposed optimizing approaches and evaluate the latency. Figure~\ref{fig:breakdown} illustrates the normalized execution time to \textbf{RPM}. Directly adopting the software scheduling on a traditional four-stage router brings 46\% execution time reduction. It balances the traffic load across the network as well as increase the channel utilization of the network. The reduced router pipeline contributes to 6\% execution time reduction for the shorter head flit traversal delay. At last, the hybrid routing and collapsing the data chunk into single packet reduce 9\% overall execution time because of the less controlling overheads. 



\subsubsection{Energy Saving:}

To demonstrate the energy efficiency of \name, we estimate the power of computation cores using Timeloop. The power of the \name~routers is estimated by Cadence Genus. We then calculate the energy consumption of both communication and computation using the product of the estimated power and total execution cycles. We test the energy improvement over a baseline unicast NoC that integrates 4 virtual channels for each port and a multiplexer-based switch. As Figure~\ref{fig:breakdown} illustrates, \name~ reduces 74\% energy and 61\% energy. The computation energy saving mainly comes from the reduced execution time. While the lower router power (benefited from the simplified hardware architecture) and reduced execution cycles jointly contribute to the low communication energy consumption.

\subsubsection{Scalability Study: }

\begin{figure}[h]
  \centering
  \includegraphics[width=0.5\textwidth]{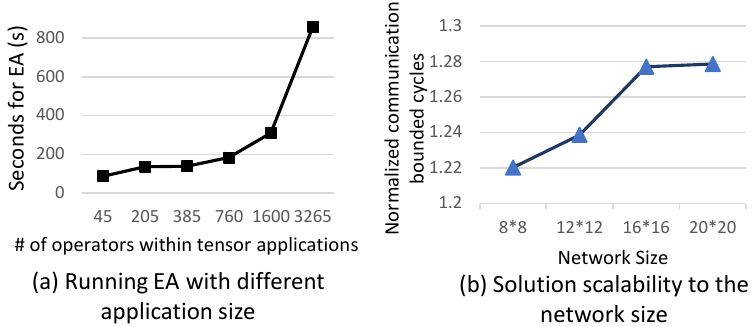}
  \caption{(a) indicates the running time with different application. (b) solution scalability to the network size. }
  \label{fig:time}
\end{figure}
\vspace{-1em}
Figure~\ref{fig:time}~(a) shows the timing consumption of METRO software scheduling framework on various sizes of tensor application graphs. The execution time of METRO framework can generally be kept within 15 minutes in our experiments, while showing a good linear growth trend with respect to the size of the incoming tasks. To illustrate the scalability of our approach, we test Mobilenet-v3 on spatial architectures with different sizes. Figure~\ref{fig:time}~(b) illustrates the solution's quality is stable with respect to the network size. We normalize the communication bounded cycles to the computing cycles, and the normalized communication bounded cycles only increase 6\% as the size of spatial architecture becomes 6.25x larger.

\section{Conclusion}
When processing tensor applications on a spatial accelerator, the unbalanced traffic may cause serious congestion and under-utilization problems in NoC.  
The fundamental reason is that a hardware NoC only makes decisions according to the nearby traffic information within a short  period  of  time. On the contrary, the software can obtain all global traffic information for processing a tensor task. Thus, we can leverage such prior information and offload the traffic scheduling from hardware fabrics to the software framework. The sophisticated routing and traffic control at the software level can handle the problems efficiently. At the same time, the hardware design of NoC can be simplified. Such a benefit is also attractive for large-scale spatial accelerators. Experimental results show that the software approach can significantly outperform traditional hardware designs,  in respect  of  performance, energy consumption, and chip area. 



\bibliographystyle{plain}
\bibliography{references}

\end{document}